# Thermal transport characteristics of impinging ferrofluid droplets in the presence of a magnetic field

Ram Krishna Shah[1, *], Saptarshi Mandal[2]

[1]Mechanical Cluster, School of Advanced Engineering, UPES Dehradun, India – 248007

[2]Department of Aerospace Engineering, Amity School of Engineering and Technology, Panvel, Maharashtra - 410206

## Abstract

Droplet–surface interactions are critical in both natural phenomena and engineered systems. While conventional droplet impingement is well understood, the behavior of nonaqueous droplets under external fields remains less explored. This study investigates the influence of magnetic fields on ferrofluid droplet impingement on heated surfaces, leveraging the super-paramagnetic properties of ferrofluids to alter impact dynamics. A multiphysics computational model incorporating magnetic field generation, fluid flow, interface tracking, and heat transfer is employed to examine the effects of Weber number and surface wettability on spreading and thermal transport. The application of a magnetic field introduces an attractive body force within the ferrofluid, significantly increasing the maximum spreading diameter (up to 35%) and suppressing rebound on hydrophobic surfaces. This leads to enhanced contact time and up to 150% improvement in heat transfer. Additionally, the magnetic field modifies the internal velocity field, resulting in steeper velocity gradients and a substantial rise in wall shear stress, an effect that further contributes to thermal enhancement. These findings highlight the magnetic field's ability to actively control droplet behavior and heat transfer, offering promising avenues for dynamic thermal management applications such as magnetically actuated thermal switches.

***Corresponding Author:** ramk.sha@ddn.upes.ac.in ; rammax8@gmail.com

## 1. Introduction

Understanding droplet impingement on solid surfaces is essential across diverse fields, from spray cooling and coating processes to agriculture and disease transmission. Recent technological advancements, particularly in electronics miniaturization and heterogeneous integration, have driven the need for efficient thermal management due to increased heat flux generation. Spray cooling, utilizing the impact of numerous droplets, offers a promising solution[1,2]. Optimizing this technique requires a fundamental understanding of individual droplet interactions with heated substrates, encompassing spreading, receding, rebound, and associated heat transfer mechanisms. [3,4]. These interactions are significantly influenced by droplet size, velocity, impact angle, and surface properties. Consequently, extensive research has focused on elucidating the impingement dynamics of aqueous droplets, leading to substantial advancements in our comprehension of these phenomena[5,6]

While extensive research has explored the impact behavior of conventional aqueous droplets, the dynamics of droplets with unique properties, such as ferrofluids or ionic fluids, remain largely unexplored. Ferrofluids, colloidal suspensions of magnetic nanoparticles in a carrier liquid, offer intriguing possibilities for manipulating droplet behaviour through external magnetic fields, leading to their applications in microfluidics [7,8], heat transfer enhancement [9–13], droplet manipulation platforms [14,15], and 3D printing [16]. Some of these applications necessitate a fundamental understanding of ferrofluid droplet interactions with substrates of varying wettability under magnetic influence. Consequently, recent investigations have begun to explore ferrofluid droplet impingement in magnetic fields. For instance, Ahmed et al. [17,18] experimentally studied the effect of magnetic field strength (produced by permanent magnets of finite size) on the maximum spreading diameter, proposing semi-empirical correlations to predict such behaviour. They observed that the maximum spreading diameter could be tuned by varying the applied magnetic field strength without altering the Weber number, while the time to reach maximum spreading remained constant. Similarly, Zhou and Jing [19] conducted experimental research on the impact dynamics of ferrofluid droplets under a vertical magnetic field. They reported a minimal effect of the magnetic field (non-uniform field created by an electromagnet) on the spreading diameter and oscillations. Hasan and Wang [20] studied the effect of vertically oriented uniform magnetic fields on the spreading dynamics of ferrofluid droplets on hydrophilic surfaces. They observed a decrease in maximum spreading diameter and an increase in droplet height with an increase in the intensity of the applied field strength. Li et al. [21] proposed a theoretical model to predict the maximum spreading diameter for a wide range of magnetic fields and Weber numbers and validated it against their experimental observations. They observed a decrease in the maximum spreading diameter till a critical Weber number and no effect thereafter, in the presence of a magnetic field. Huang et al. [22] developed a similar approach to predict the maximum spreading diameter and validated their model against their numerical results. Benther et al. [23] conducted an experimental investigation to examine the effect of a magnetic field on impingement dynamics and the associated heat transfer characteristics of ferrofluid droplets. They observed significant enhancement (up to 2.5 times) in the heat transfer rate, especially for hydrophobic surfaces in the presence of the magnetic field.

It is evident from the literature review that a more comprehensive understanding is required regarding the influence of magnetic fields, especially for the non-uniform fields produced by realistic finite-size magnetic sources, on the maximum spreading of droplets across both hydrophobic and hydrophilic surfaces. Moreover, the crucial aspect of associated heat transfer has been addressed in only a single recent experimental study, highlighting a significant gap. Numerical modelling of these coupled phenomena promises valuable insights for numerous applications. To address this, the subsequent paragraphs present an in-depth literature review of the Phase-field (PF) methodology, the chosen approach for modelling contact line dynamics and fluid-fluid interface evolution in this work.

The Phase-field method is a diffuse interface (DI) method based on the free energy of the physical systems. It is a mathematical technique used to define and capture the dynamic evolution of the interfaces between multiple phases of physical systems. PF methods were originally developed to capture phase transition phenomena, such as nucleation, evaporation, and solidification. More recently, PF methods have been used in diverse fields of applications such as phase transformation, solidification dynamics, fracture mechanics, multiphase flow, etc., for capturing the dynamic evolution of the interfaces and topological changes in multiphase systems [24,25]. In the context of binary fluid systems, PF methods represent a more physical approach to capturing the dynamics of the interfaces between the two fluids than other popular models, such as Volume of Fluid and Level Set methods, because of their free energy approach. The wetting and no-slip boundary conditions for moving contact line (MCL) problems can be easily implemented in PF modelling, unlike in Volume of Fluid and Level Set methods, where a workaround is required to avoid stress singularities at contact line motion [25].

Jacqmin first implemented the Cahn-Hilliard Phase-field (CH PF) model for the modelling of two-phase flows and moving contact line (MCL) problems [26,27]. He showed that such a model can be successfully implemented in modelling interfacial flows. He also reported that the PF model's mobility parameter ($\gamma$) is a material-dependent property, and too large or too small a mobility would produce unrealistic results. A scaling relation for the mobility with the capillary width or interfacial thickness ($\varepsilon$) as $\gamma \sim \varepsilon^2$ was reported in his work. As $\varepsilon \rightarrow 0$, the diffuse interface in PF approaches the classical sharp-interface limit. Yue and co-workers extensively investigated PF methods for the modelling of interfacial dynamics in the two-phase immiscible flow of complex/non-Newtonian fluids and MCL problems [28–30]. Apart from interfacial dynamics, the PF model was also tested for the modelling of the complex rheology of fluids in their works. It was reported that the variational framework-based PF models could easily accommodate the modelling of complex fluid behaviors. In their work on the MCL problems, Yue suggested that the mobility and capillary width are independent of each other as $\gamma \sim \varepsilon^0$ and mobility need not be compensated for the change in capillary width ($\varepsilon$), in contrast to the suggestions of Jacqmin [30]. They also pointed out that a sharp-interface limit can be approached by scaling mobility with capillary width as $\gamma \sim \varepsilon^2$. Ding et al. employed PF modelling to investigate interfacial dynamics of density contrast fluids (e.g. gas and liquid) [31]. The investigation by Yue and co-workers mentioned above has mostly focused on the density-matching fluids because they observed non-convergence for the density contrast fluids. Ding suggested a scaling of mobility in the form of Peclet number (Pe) with the Cahn number (Ca) as $\mathrm{Pe} \sim (\mathrm{Ca})^{-2}$ in the PF modelling of density contrast fluids. Cai and co-workers have worked extensively on the PF-based numerical modelling of the immiscible two-phase flows of gas and liquid in different interfacial flow applications [32,33]. They implemented a PF model based on the CH-PF equation in the *OpenFOAM*® (called *PhaseFieldFoam*) and tested its applicability in MCL and other two-phase flow problems. They have also performed a detailed investigation on the effect of PF model parameters such as mobility, interfacial resolution, and thickness on the numerical results in the context of MCL/two-phase flow problems of gas-liquid flows, such as spreading of drops or formation of bubbles. Several suggestions made by earlier investigators regarding the tuning of the model parameters were tested by their group. They reported that the resolution of capillary width or interfacial thickness (DI) by four or higher mesh elements for $\varepsilon \leq 0.02L_c$ (where $L_c$ is the characteristic length) would provide numerical results independent of $\varepsilon$. Their suggestions for the selection of PF parameters have also been implemented in the present work. In recent work, Bai et al. [34] conducted a numerical investigation on the droplet formation in a liquid-liquid immiscible two-phase flow in a microfluidic flow-focusing geometry using the PF method. They implemented the selection criteria for the mobility parameter suggested by Yue et al. [35] and examined the effect of different orders of mobility parameters on the size and shape of the droplet formation. The simulation results were compared with similar experimental observations, and it was reported that PF mobility is a phenomenological parameter whose magnitude can be accurately determined by comparing the simulation results with experimental observations. In a recent work, Samkhaniani et al. [36] employed the PF model, *PhaseFieldFoam*, developed by Cai et al. earlier, to model droplet impingement dynamics on heated hydrophobic surfaces. They observed that the PF model can adequately resolve the interfacial evolution during impact dynamics. Their simulation results closely match the experimental observations reported by Guo et al. [37]. They also reported that the wetting characteristics did not significantly affect heat transfer for the investigated range of contact angles.

It is evident from the literature that the PF methods are capable of modelling complex phenomena such as MCL problems and topological changes in two-phase immiscible and other interfacial flows. Their free energy-based modelling approach and thermodynamic consistency make them suitable for the modelling of interfacial dynamics in a wide variety of physical problems. The no-slip boundary condition, which is often challenging to implement in sharp-interface methods due to the occurrence of stress singularity, can be implemented in the PF method. However, it can also be inferred from the earlier studies that the tuning of PF model parameters is required for accurate numerical results.

In the present work, PF-based computational modelling of the impingement dynamics of ferrofluid droplets is carried out on a solid substrate across a range of Weber numbers and wettability characteristics, both in the absence and presence of a magnetic field. Firstly, a validation exercise has also been carried out to establish the accuracy of the obtained computational results. The validation study is further extended to examine the impingement dynamics of aqueous droplets and compare results with previous works. In the second part, the investigation delves into the impact dynamics of ferrofluid droplets on heated surfaces under the influence of an applied magnetic field, shedding light on the interplay between magnetic forces, fluid flow, and heat transfer during droplet impact.

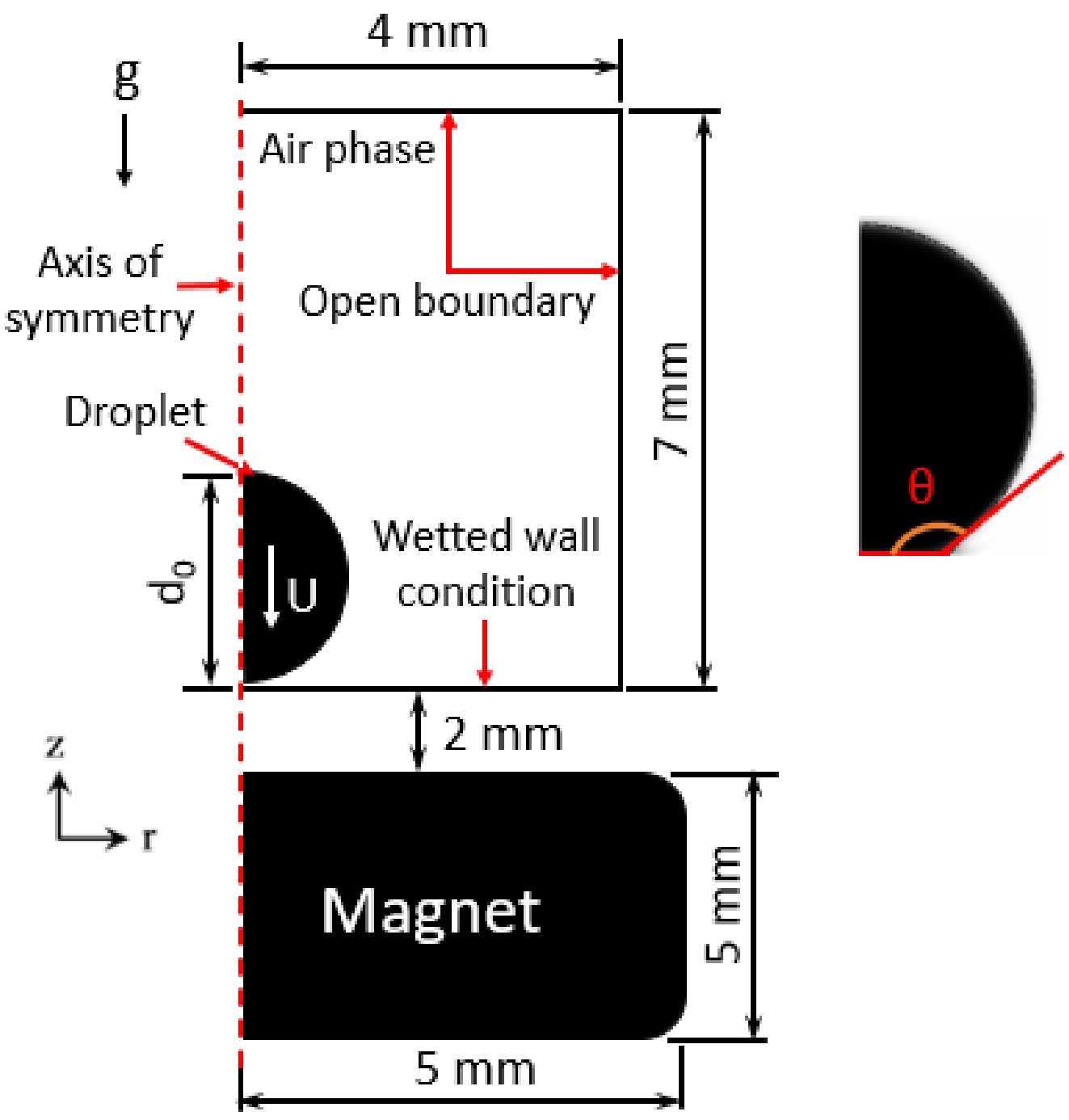

***Figure 1:*** *Schematic of the problem statement with dimensions and boundary conditions*

## 2. Computational Modelling and Methodology

The impingement dynamics of ferrofluid droplets are modelled on a smooth solid heated substrate in the presence of an externally applied magnetic field (MF) at different Weber numbers (We) and wetting conditions (equilibrium contact angles, $\theta_e$), spanning both hydrophilic and hydrophobic regimes. The applied magnetic field is produced by simulating a finite-size permanent magnet. The Weber number ($We = \rho U^2 d_0/\sigma$), representing the ratio of kinetic to surface energy, assesses the relative influence of inertial and surface tension forces during morphological evolution. At high We, inertial forces dominate, leading to droplet breakup and wave formation, while at low We, surface tension prevails, resulting in stable droplet formation. The Reynolds number ($Re = \rho U d_0/\mu$), the ratio of inertial to viscous forces, and the magnetic Bond number ($Bo_m = B^2 d_0/\mu_0\sigma$), the ratio of magnetic to surface tension forces, is employed to analyze the effects of the relative importance of different forces during droplet dynamics. Here, $\rho$, $\mu$, and $\sigma$ represent the density, viscosity, and surface tension of fluids, respectively; $U$ is the initial droplet impact velocity; $d_0$ is the initial droplet diameter; and $B$ is the applied magnetic field intensity in Tesla. The magnitude of magnetic permeability, $\mu_0 = 4\pi \times 10^{-7}\ N/A^2$.

A single set of continuity, momentum and energy equations is solved numerically to simulate the thermal transport characteristics of the impinging droplet. The multiphase flow is modelled as incompressible, immiscible, and non-boiling flow, and ferrofluid (*ff*) is modelled as a pseudo-single-phase Newtonian liquid with magnetic characteristics. The time scale of the impinging dynamics is of the order ~ O ($10^{-3}$ s), which is much larger than the relaxation time scale for material magnetization ~ O ($10^{-5}$ s). Therefore, it is assumed that the magnetization of MNPs in suspension will remain aligned with the direction of applied MF, and thus, angular momentum considerations have been neglected in the momentum equation. The interfacial evolution of impinging droplet dynamics (spreading, receding and bounce-off phases) is simulated using the Phase-field (PF) formulation, a diffuse interface-based methodology for modelling interfacial multiphase flows [27]. In the PF method, a scalar ordered parameter called the phase-field parameter ($\phi$) is used to differentiate between the two phases. It is assigned two distinct values (-1 for air and +1 for ferrofluid) in the bulk of the material phases, with a rapid but smooth transition in the interfacial regions.

An axis-symmetric schematic of the problem statement with relevant dimensions and boundary conditions is shown in Figure 1. This same setup is modelled and simulated to study the droplets' impingement dynamics and thermal transport characteristics in the absence and presence of a magnetic field. A cylindrical permanent magnet of dimensions 5 mm (thickness) × 10 mm (diameter) is numerically modelled using the Gauss divergence equation (Eq. 1) to simulate the applied MF. The magnet is placed 2 mm below the air domain, as shown in Figure 1. An air domain is defined around all the objects to simulate the magnetic field. The following equations with specified boundary conditions are numerically solved to create the applied MF,

$$\boldsymbol{\nabla}.\boldsymbol{B} = 0; \boldsymbol{B} = \mu_0(\boldsymbol{H} + \boldsymbol{M}) \tag{1}$$

where $\boldsymbol{B}$, $\boldsymbol{H}$ and $\boldsymbol{M}$ represent magnetic induction (or flux density, in Tesla [T]), magnetic field (A/m) and magnetization (A/m), respectively. The above equations are subjected to magnetic insulation boundary conditions, $n.\boldsymbol{B} = 0$ at the outer domain walls ($n$ represents the normal unit vector at the boundary) [9]. The magnetic characteristics (magnetization, $\boldsymbol{M}$) of ferrofluid droplets are defined using the Langevin function [13,38] as,

$$\boldsymbol{M} = \boldsymbol{M_S} \boldsymbol{L}(\boldsymbol{\alpha}) \frac{\boldsymbol{H}}{|\boldsymbol{H}|}; \boldsymbol{L}(\boldsymbol{\alpha}) = \mathbf{cot}\,\boldsymbol{h}\,(\boldsymbol{\alpha}) - \left(\frac{\mathbf{1}}{\boldsymbol{\alpha}}\right); \ \boldsymbol{\alpha} = \frac{\boldsymbol{\mu_0 M_d H d_h}^3}{\boldsymbol{6 k_B T}} \tag{2}$$

The Langevin function (Eq. 2) accurately depicts the non-linear magnetization characteristics shown by ferrofluids [13]. Here, $L(\alpha)$ denotes the Langevin function, with $\alpha$ being the ratio of magnetic energy to thermal energy. The equation clearly indicates that the magnetization ($\boldsymbol{M}$) of ferrofluids is influenced by various factors, including the fluid temperature ($T$), the size ($d_h$) and concentration ($\varphi$) of the magnetic nanoparticles (MNPs), the material's domain magnetization ($M_d$), and the applied magnetic field ($\boldsymbol{H}$). Furthermore, the saturation magnetization ($M_s$) of ferrofluids is determined by the domain magnetization ($M_d$) and the volume fraction ($\varphi$), following the relation, $M_s = \varphi M_d$. The magnetic body force ($\boldsymbol{f_k}$) induced in ferrofluid in the presence of a magnetic field is defined as follows,

$$\boldsymbol{f_k} = \boldsymbol{\mu_o}(\boldsymbol{M}.\boldsymbol{\nabla})\boldsymbol{H} \tag{3}$$

The induced magnetic force can be varied either by changing the gradient of the applied magnetic field or by magnetization ($M_s$) of ferrofluids, as seen in Eq. 3.

The transient multiphase flow phenomena and the interfacial evolution are simulated using continuity equation (Eq .4), Navier Stokes (N-S) equations (Eq. 5), and Cahn-Hilliard Phase-field (CH PF) equations (Eq. 7). The thermal transport between substrate and droplet is modelled using the energy equation (Eq. 6). The solved continuity, momentum (N-S) and energy equations are as follows,

$$\nabla.\boldsymbol{V} = 0 \tag{4}$$

$$\frac{\partial \boldsymbol{V}}{\partial t} + \rho(\boldsymbol{V}.\boldsymbol{\nabla})\boldsymbol{V} = -\boldsymbol{\nabla}P + \mu\nabla^2\boldsymbol{V} + \boldsymbol{f}_g + \boldsymbol{f_{ST}} + \boldsymbol{f_m} \tag{5}$$

$$\frac{\partial T}{\partial t} + \boldsymbol{V}.\boldsymbol{\nabla}T = \left(\frac{k}{\rho C_p}\right)\boldsymbol{\nabla}^2 T \tag{6}$$

Where $\boldsymbol{V}$ and $P$ are velocity (in vector form) and pressure terms, respectively. The $\boldsymbol{f}_g$, $\boldsymbol{f}_{ST}$ and $\boldsymbol{f}_m$ terms in momentum equations represent gravity, surface tension and magnetic forces, respectively, which are defined in subsequent sections. The $\boldsymbol{f}_g$ term is defined as the product of fluid density and gravitational constant, $g$.

The above equations are subject to the following boundary conditions. The no-slip ($\boldsymbol{V}$ = 0) boundary condition is defined for the bottom surface, and zero-gauge pressure ($P$ = 0) is defined for the side and top walls. A uniform and constant wall temperature ($T$ = 60°C) boundary condition is specified on the bottom wall of the domain. The outflow boundary ($\boldsymbol{n}.\,\boldsymbol{q''}$ = 0, where $q''$ is heat flux defined as $q'' = -k\nabla T$) condition is defined for the top and side walls of the domain shown in Figure 1. The initial conditions for velocity and temperature are $\boldsymbol{V}$ = 0 and $T$ = 20°C.

The Cahn-Hilliard Phase-field (CH PF) equation, used to model interfacial dynamics, is a fourth-order convection-diffusion equation in phase field variables, $\phi$

$$\frac{\partial\phi}{\partial t} + \boldsymbol{V}\cdot\nabla\phi = \nabla\cdot\frac{\gamma\lambda}{\varepsilon^2}\nabla\Psi;\ \Psi = -\nabla\cdot\varepsilon^2\nabla\phi + (\phi^2-1)\phi \tag{7}$$

Where $\boldsymbol{V}$ is fluid velocity and $\gamma$, $\lambda$, and $\varepsilon$ are called the mobility, mixing energy density and interface thickness parameters, respectively. Parameter $\phi$ takes a value of -1 and +1 in the bulk of the binary fluid system (gas and liquid in the present case) with a rapid but smooth transition in the interfacial region. The interface between two fluids is defined by $\phi = 0$ contours. The phase-field mobility ($\gamma$) is another critical parameter defined in the CH-PF equation. It governs the time scales of advection and diffusion terms of the phase-field parameter ($\phi$) and affects an out-of-equilibrium interface's time scale of relaxation [34]. A scaling relationship between the mobility and interface thickness parameters as $\gamma \sim \varepsilon^2$ is suggested by Jacqmin [27] for a stable evolution of the interface, and this is also implemented in the PF formulation on the COMSOL Multiphysics® (V 5.3a) platform. In the present simulations, $\gamma$ is defined as $\gamma = \chi\,\varepsilon^2$, and its magnitude is controlled through a mobility tuning parameter ($\chi$).

The variable $\psi$ (called the phase-field help variable) is used to divide the fourth-order CH PF equation into two second-order PDEs, and ε, as its name suggests, controls the thickness of the interface. The relationship between surface/interfacial tension ($\sigma$), $\lambda$ and $\varepsilon$ is given by

$$\sigma = \frac{2\sqrt{2}}{3}\frac{\lambda}{\varepsilon} \tag{8}$$

The chemical potential ($G$) of the two-fluid system and surface tension force ($\boldsymbol{f}_{ST}$) are related as,

$$G = \frac{\lambda\Psi}{\varepsilon^2};\ \boldsymbol{f}_{ST} = G\,\nabla\phi \tag{9}$$

The volume fraction of individual fluids and the thermophysical properties of the two-phase mixture (air and liquid) used are defined in terms of $\phi$ as,

$$Vf_1 = \frac{1-\phi}{2};\ Vf_2 = \frac{1+\phi}{2} \tag{10}$$

$$\rho_{\mathrm{TP}} = \rho_{\mathrm{air}}Vf_1 + \rho_{\mathrm{ff}}Vf_2;\ \mu_{\mathrm{TP}} = \mu_{\mathrm{air}}Vf_1 + \mu_{\mathrm{ff}}Vf_2 \tag{11}$$

$$k_{\mathrm{TP}} = k_{\mathrm{air}}Vf_1 + k_{\mathrm{ff}}Vf_2;\ C_{p_{\mathrm{TP}}} = C_{p_{\mathrm{air}}}Vf_1 + C_{p_{\mathrm{ff}}}Vf_2$$

Where $Vf_1$ *and* $Vf_2$ represent the volume fraction of gas and liquid phases. The subscripts *air*, *ff* and *TP* are used for the gas (air), ferrofluid, and two-phase mixture phases. The symbols $\rho$, $\mu$, $k$, and $C_p$ represent density, viscosity, thermal conductivity and specific heat (thermophysical properties) of individual and mixture phases. The following properties of the fluids are used in simulations:

**Table 1:** Fluid properties used in the computational modelling

| Material | Density ($\rho$, kg/m$^3$) | Viscosity ($\mu$, Pa-s) × $10^{-5}$ | Thermal Conductivity ($k$, W/m-K) | Specific heat ($C_p$, J/kg-K) | Surface tension ($\sigma$, N/m) |
|---|---|---|---|---|---|
| Air | 1.3 | 1.5 | 0.026 | 1000 | - |
| Water | 998 | 90 | 0.60 | 4180 | 0.070 |
| Ferrofluid | 1018 | 91 | 0.60 | 4090 | 0.035 |

## 3. Results and discussion

### 3.1 Validation

Firstly, a validation study is performed to assess the accuracy of computational modelling of such multiphase flow phenomena using the Phase-field formulation implemented on the COMSOL Multiphysics platform. The experimental work of Guo et al. [37] and the computational modelling of the same work by Samkhaniani et al. [36] are considered for validation. They implemented *PhaseFieldFoam* on the OpenFOAM platform using their numerical modelling approach and investigated the thermal transport characteristics of impinging aqueous droplets on heated hydrophobic/superhydrophobic substrates. Figure 2(a) presents a comparison of the present work with Guo and Samkhaniani's works for the temporal interfacial evolution of the droplet after impact for Weber number (We) = 20, $d_0$ = 2.3 mm and contact angle ($\theta_e$) of 120°. A quantitative comparison of the normalized maximum spreading diameter called the spreading parameter, $\beta(t) = d(t)/d_0$, is also presented in Figure 2(b). As seen in the referenced figures, the present simulation results are in good agreement with the previous works. The maximum spreading diameter during advancing and receding phases closely matches experimental and numerical observations reported in earlier studies. This validation exercise provided the necessary confidence in the present modelling approach. The present simulations are carried out at Cahn number (Cn) = 0.01, resulting in a maximum mesh element size ($h_{max}$) of $2.3\times10^{-5}$ m or 23 µm (Cn is defined as the ratio of maximum mesh element size to initial droplet diameter, Cn = $h_{max}/d_0$). The Cn = 0.01 provides four mesh elements in the interfacial regions for the resolution of the interfacial thickness (defined by contours of $-0.9 < \varphi < 0.9$), which is generally considered sufficient to resolve the interfacial region [33,34]. This same mesh size (corresponding to Cn = 0.01) is used for all the studied cases in the present work. The variation of PF parameter $\phi$ in the range -0.9 to +0.9 in the interfacial region represents 98.5% of the interfacial stress and is generally considered as the thickness of the interface [39,40]. The value of the mobility tuning parameter ($\chi$) is kept at 1 in all the simulations.

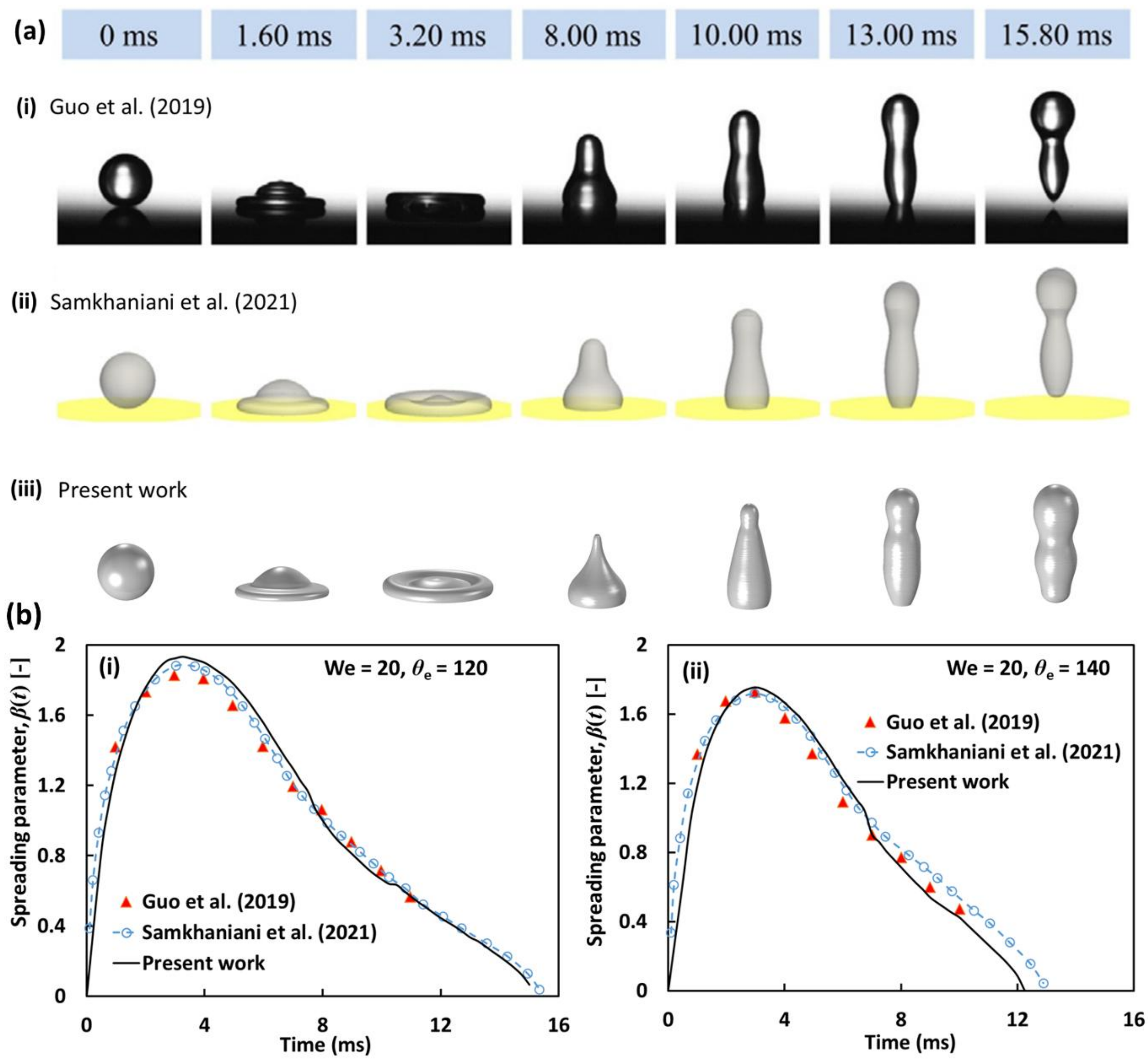

**Figure 2:** *(a) Comparison of present simulations results for temporal interfacial evolution of the impinging droplet with earlier works for We = 20 and $\theta_e$ = 120° (b) Comparison of results for normalized spreading parameter, $\beta(t) = d(t)/d_0$ with previous studies at We = 20 and (i) $\theta_e$ = 120° (ii) $\theta_e$ = 140°.*

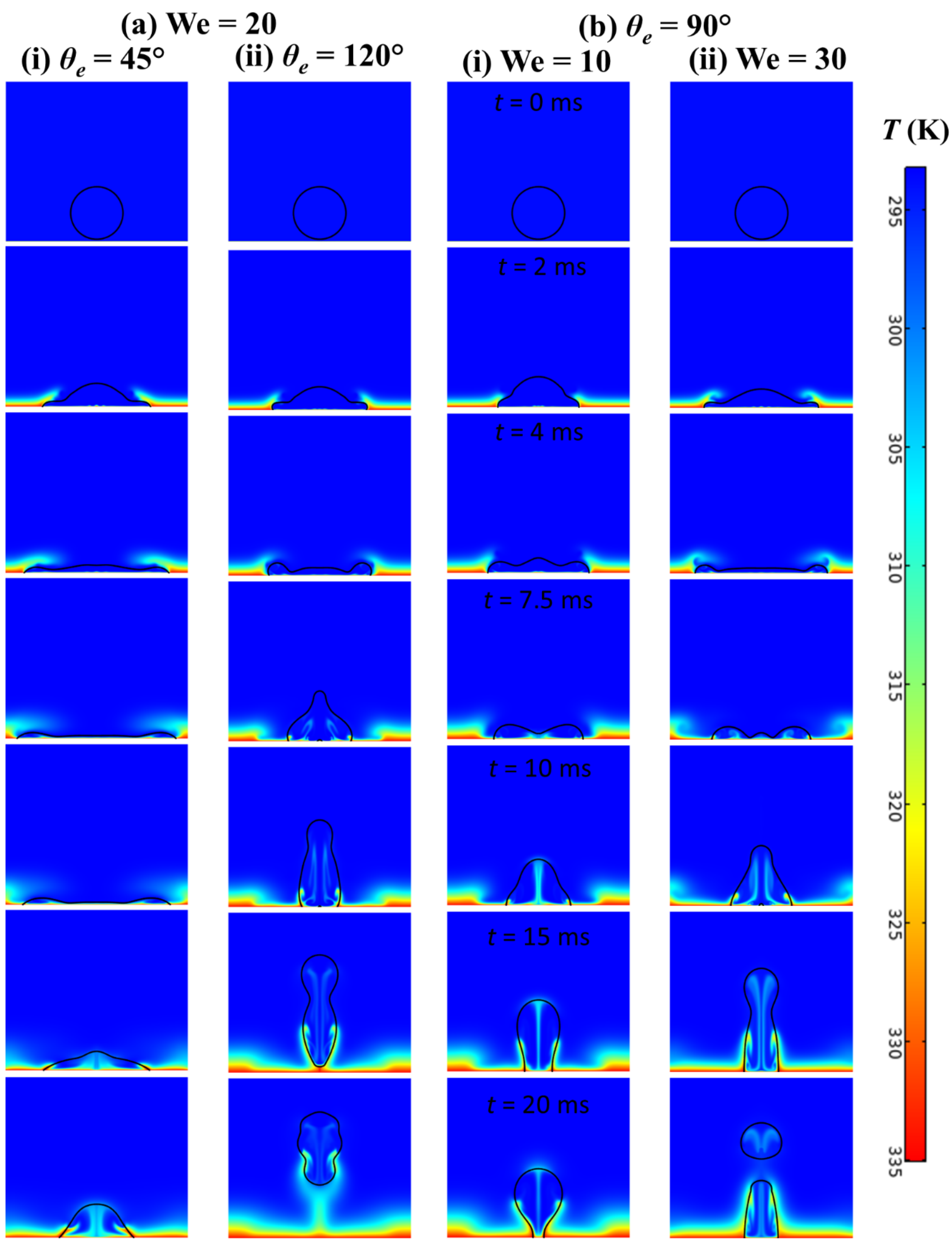

***Figure 3:*** *A comparison of morphological evolutions (advancing and receding phases) of the droplet after impact at (a) We = 20 for (i) $\theta_e$ = 45° and (ii) $\theta_e$ = 120° and at (b) $\theta_e$ = 90° for (i) We = 10 and (ii) We = 30 cases.*

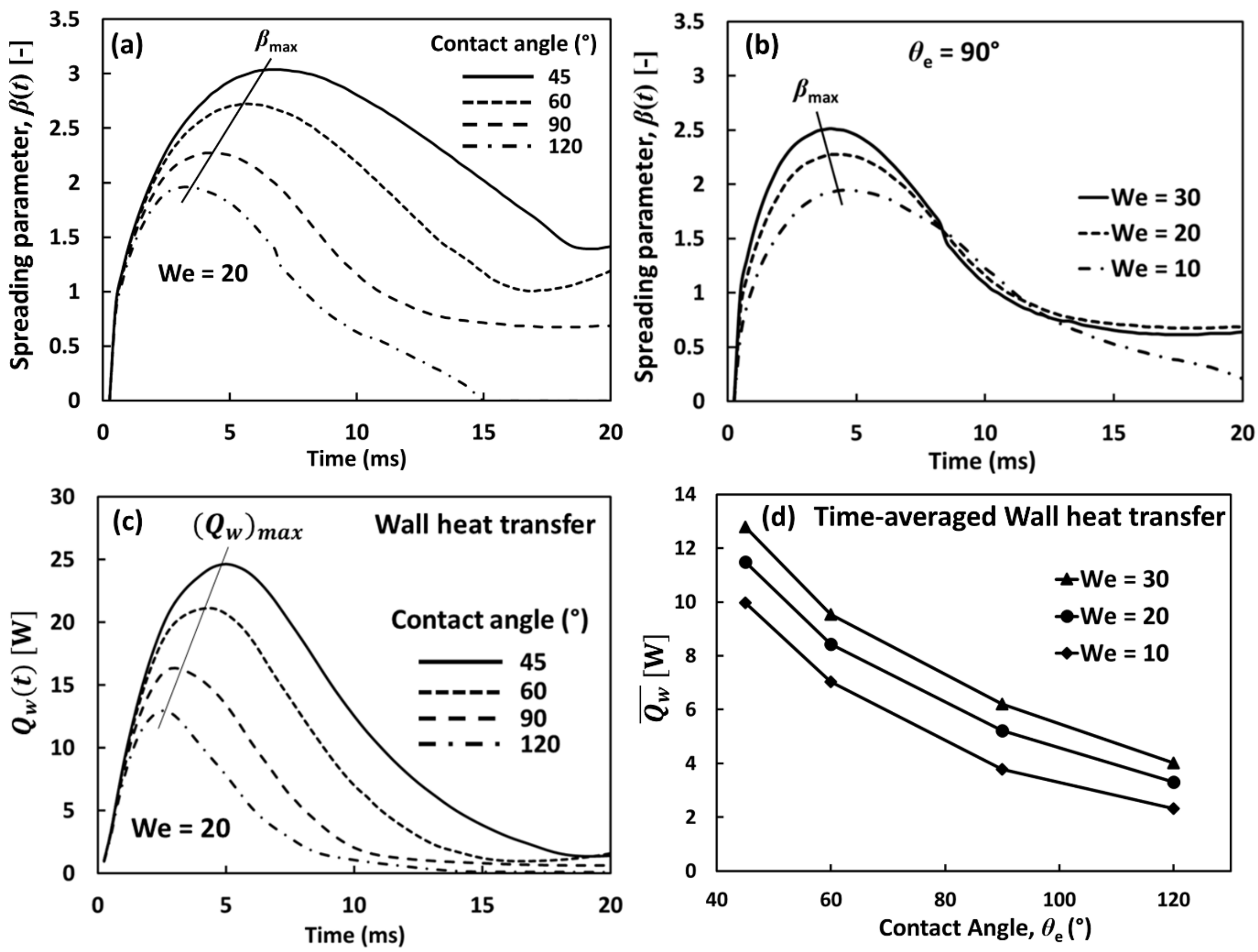

***Figure 4:*** *Spreading parameter, β(t) during interfacial evolution for (a) different wetting conditions (contact angles) at We = 20 and (b) different Weber numbers at $\theta_e$ = 90° (c) associated temporal variation of total wall heat exchanged with the evolution of the impinging droplet for different contact angles at We = 20 (d) Time-averaged (20 ms) wall heat transfer for different Weber numbers and contact angles.*

## 3.2 Aqueous droplets (No-field cases)

The validation study is extended to investigate the effect of inertia force (through Weber number) and wetting characteristics (equilibrium contact angle, $\theta_e$) on the spreading parameter, $\beta$, and associated heat transfer characteristics. The Weber number is varied in the range of 10 to 30, and the contact angle in the range of 45° to 120°, covering both hydrophilic and hydrophobic regimes. The corresponding impact velocity Reynolds number is in the range of 1022 – 3066. Figure 3(a, b) shows a comparison of the morphological evolution of the droplet after impact and associated temperature distribution for hydrophilic ($\theta_e$ = 45°) and hydrophobic ($\theta_e$ = 120°) cases at We = 20. A distinction between advancing and receding phases can be seen clearly in both cases. As seen from the figure, no bounce-off is observed during the receding phase for the hydrophilic surface. The initial position ($t$ = 0) of the droplet can also be seen in the snapshots of Figure 3.

Across the investigated Weber number range, both inertia and surface tension significantly influenced droplet impact dynamics. The inertia drives the initial spreading, converting kinetic energy to surface energy as the contact area increases. Hydrophilic surfaces promoted spreading due to strong fluid-surface attraction, while hydrophobic surfaces hindered it. As the droplet reached maximum spread, inertial effects diminished, and capillary and viscous forces became dominant. On hydrophilic surfaces, strong surface attraction resisted retraction, leading to a slower receding phase, a larger final contact area, and, consequently, higher contact time. Conversely, weak attraction on hydrophobic surfaces resulted in rapid retraction and recoil, often with complete or partial rebound. This difference in spreading and receding behavior significantly impacted heat transfer. The hydrophilic surface ($\theta_e = 45°$) exhibited the highest contact time and, consequently, the highest heat transfer (Figure 3(d) and 4(d)). Heat transfer for the $\theta_e = 45°$ cases was approximately two to three times greater than the hydrophobic ($\theta_e = 120°$) cases, depending on the Weber number. Figure 4(a-c) illustrates the effect of the Weber number on spreading and receding. The maximum droplet deformation occurs around 3 to 6 ms, depending on the surface wettability and Weber number. The maximum spreading diameter increased with the Weber number, as expected, due to higher impact inertia. Consequently, the highest heat transfer was observed at We = 30, approximately 30% higher than the We = 10 case for $\theta_e = 45°$. Figure 4(d) summarizes the overall heat transfer for all cases. The total wall heat transfer was computed for We = 10 − 30 and $\theta_e = 45° - 120°$. The droplet impact time scale is on the order of 10 − 20 ms (Figures 2, 3 and 4), and the time-averaged heat transfer (Figure 4(d)) is computed by averaging transient heat transfer data over 20 ms time scale. As expected, the maximum heat transfer occurred at the highest Weber number (We = 30) and lowest contact angle ($\theta_e = 45°$). A high Weber number signifies high droplet inertia at a given surface tension. Combined with the hydrophilic regime ($\theta_e = 45°$), which maximizes interaction time with the substrate (~ 20 ms for a complete spreading and receding cycle, as seen in Figure 3(a)), this resulted in the highest overall heat transfer. Several correlations have already been proposed to predict the maximum spreading diameter and heat transfer rate during droplet impingement dynamics. The maximum spreading diameter and wall heat transfer are also highest for $\theta_e = 45°$ among all investigated cases, as seen in Figure 4(c, d).

## 3.3 Effect of magnetic field

Finally, the study investigated the influence of a magnetic field on the impingement dynamics of ferrofluid droplets. The magnet, positioned 2 mm below the fluid domain's bottom wall (Figure 1), generated a spatially varying magnetic field with a maximum intensity ($B_{max}$) of 0.23 T. Inset of Figure 7(a) presents a distribution of the magnetic field produced by the magnet. The corresponding magnetic Bond number $(Bo_m)_{max}$ is 2768. More details on modelling and validation of the magnetic field produced by the considered finite-sized permanent magnet are presented in our previous works [8,9,13]. This field induced a substantial magnetic body force (order of magnitude ~ $10^5$, force distribution shown in inset of Figure 7(b)) within the fluid domain, which, acting in the same direction, is comparable to the droplet's maximum inertial force (order of magnitude ~ $10^5$), thus synergistically enhancing the impingement dynamics and spreading characteristics, particularly on hydrophobic surfaces.

A qualitative and quantitative comparison of maximum spreading diameter, temperature distribution, and average wall heat transfer with and without the magnetic field is presented in Figures 5 and 6(a). In the absence of a magnetic field, droplets on hydrophobic and superhydrophobic surfaces ($\theta_e \geq 120°$) exhibit pronounced rebound behavior, detaching from the surface at approximately 15 ms (Figure 5(a) and Figure 6, We = 20). The application of a magnetic field effectively suppresses this rebound by enhancing inertial spreading and counteracting recoil forces. This leads to a ~ 25% increase in maximum spreading diameter and extended contact time with the heated substrate, as shown in the figures. Consequently, a substantial improvement in heat transfer is observed under magnetic field conditions. Specifically, the time-averaged heat flux increases by ~ 66% compared to the no-field case (Figure 6(a)), highlighting the potential of magnetic forces to manipulate droplet behavior and enhance thermal performance during impingement.

This enhancement is further attributed to the significant increase in wall shear stress, driven by modifications in the velocity field under magnetic influence. The magnetic body force accelerates internal flow, resulting in steeper velocity gradients near the solid surface and thereby increasing wall shear stress. Figures 7(a) and 7(b) compare the temporal variation of wall shear stress ($\tau$) for magnetic and non-magnetic cases at $\theta_e = 120°$, while Figure 6(b) presents the average wall shear stress for hydrophilic ($\theta_e = 45°$) and hydrophobic ($\theta_e = 120°$) surfaces, along with the net change induced by the magnetic field. At We = 10, both maximum and average wall shear stress values are approximately 150% and 200% higher, respectively, than those observed in the absence of a magnetic field for the hydrophobic surface. This enhancement diminishes with increasing Weber number due to the growing influence of inertial forces, which are of the same order of magnitude as the magnetic force. At We = 30, the increase in average $\tau$ reduces to ~ 75% for $\theta_e$ = 120° and ~ 50% for $\theta_e$ = 45°. Results for other contact angles fall between the two extremes represented by the hydrophilic and hydrophobic cases. Image snapshots (I, II and III) of Figure 7 also present a comparison of the velocity field at maximum shear stress for both cases. The modifications in the velocity field induced by the magnetic force can be seen in the images. The maximum velocity in the fluid domain is higher for the magnetic field case due to the additional force in the same direction.

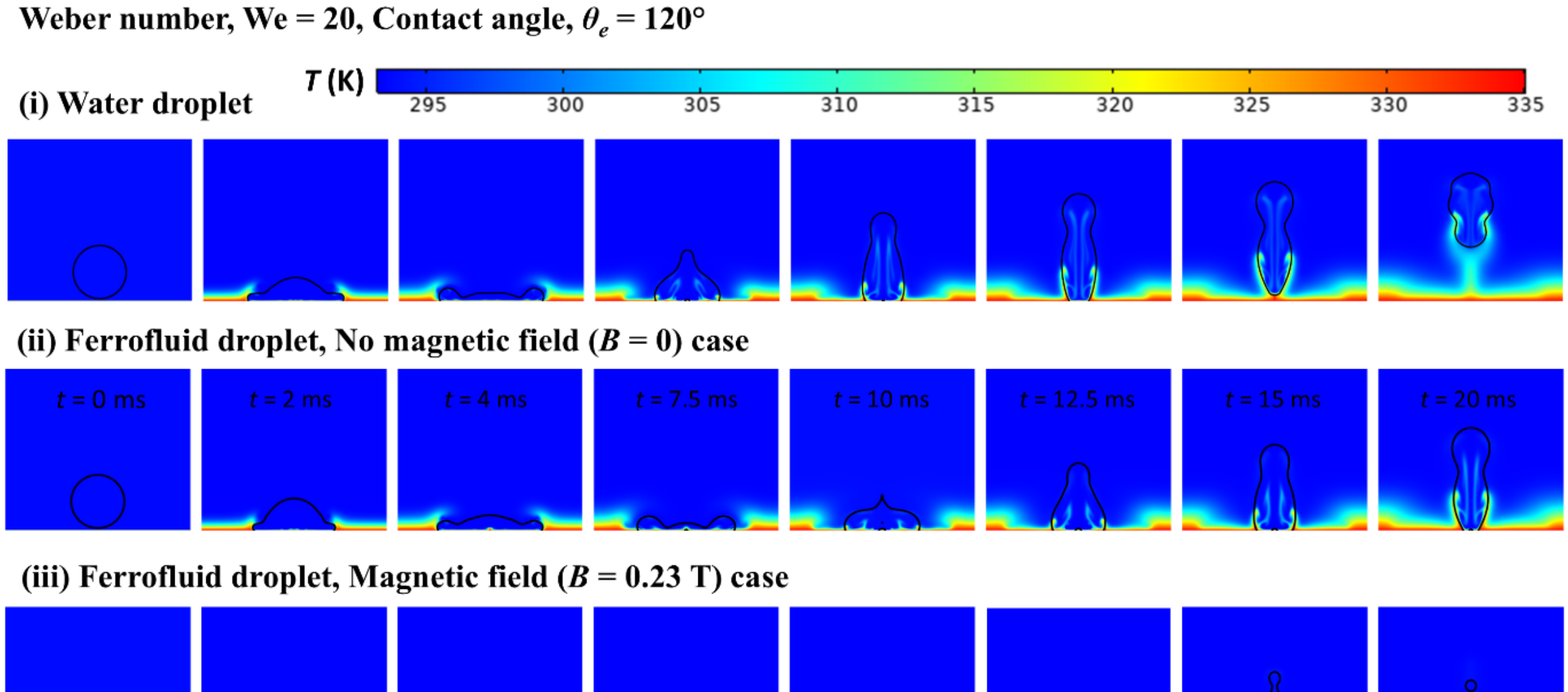

***Figure 5:*** *A comparison of droplet impingement dynamics at We = 20, $\theta_e$ = 120°) for (i) water droplet (ii) No-magnetic field cases, B = 0 (spreading, receding and rebounding phases can be seen), (iii) Magnetic field ($B_{max}$ = 0.23 T) case, again spreading and receding phases can be seen while rebounding phases have been suppressed by the magnetic force. The associated temperature distribution and the initial position of the droplet can also be seen in both cases.*

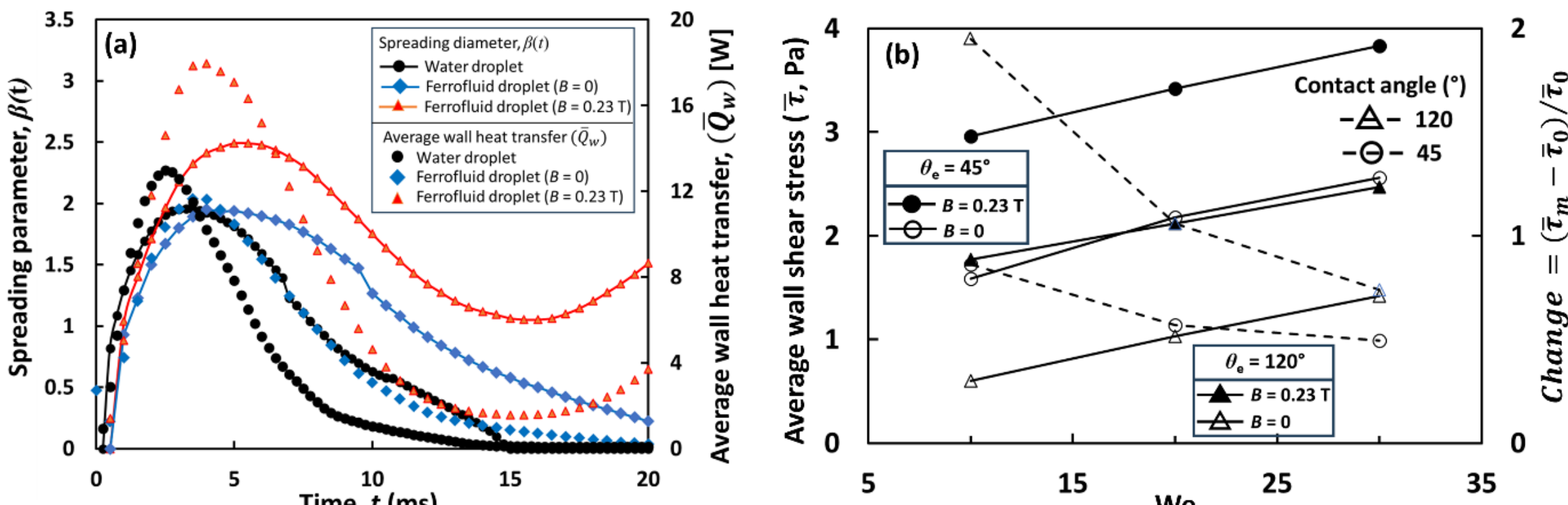

***Figure 6:*** *(a) Quantitative analysis of β(t) and average wall heat transfer ($\bar{Q}_w$) for the three cases shown in Figure 5. As seen from the plots, higher heat transfer occurs in the magnetic field case as droplet bounce-off is suppressed by the magnetic force, resulting in higher contact time between the droplet and substrate. (b) A comparison of the average wall shear stress ($\bar{\tau}$) for hydrophilic ($\theta_e$ = 45°) and hydrophobic ($\theta_e$ = 120°) cases, along with the net change induced by the magnetic field.*

The combined influence of Weber number and contact angle on spreading dynamics and heat transfer is further examined under magnetic field conditions. Image snapshots in Figure 8(a) illustrate the morphological evolution and corresponding temperature distribution for $\theta_e = 45°$ and 120° at We = 30, while Figure 8(b) presents similar data for We = 20 and $\theta_e = 90°$. Figure 9(a) shows the variation of the spreading parameter $\beta(t)$ across different contact angles at We = 20, and Figure 9(b) presents the same across various Weber numbers for $\theta_e = 90°$. As evident from Figure 8(a), the contact angle significantly influences droplet morphology even in the presence of a magnetic field. This is further supported by the $\beta(t)$ plots in Figure 9(a), where the maximum spreading parameter is highest for the hydrophilic case ($\theta_e = 45°$). The magnetic force, acting in the same direction as the inertial force, enhances spreading and increases droplet contact time, thereby improving overall heat transfer, as seen in Figure 8(b). Notably, the magnetic field contributes to a substantial improvement in heat transfer even for hydrophilic surfaces, which is also evident from the wall shear stress plots.

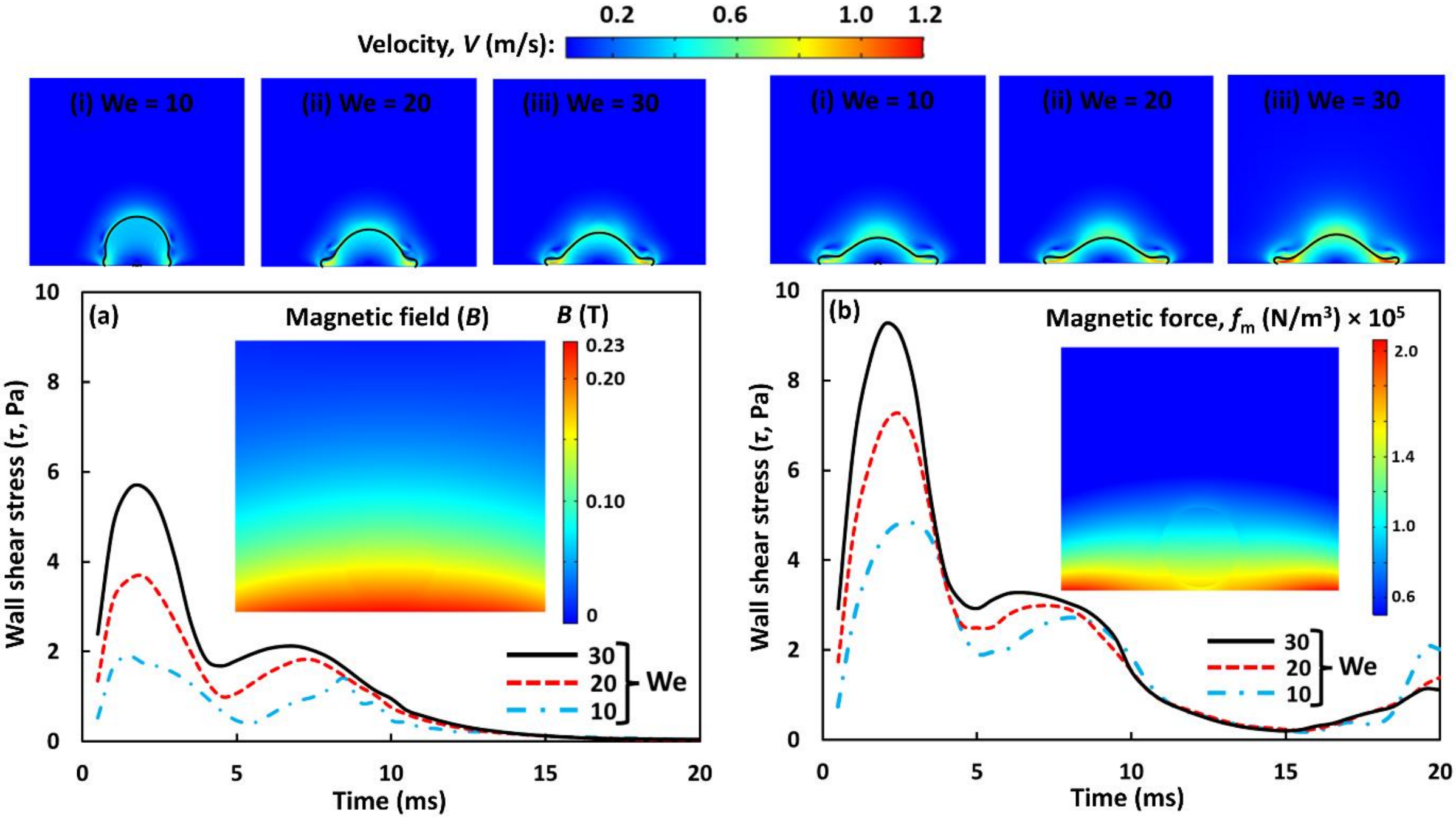

**Figure 7:** *Comparison of wall shear stress (τ) for (a) no field (B = 0) case with (b) magnetic field (B = 0.23 T) case for $\theta_e$ = 120°. Inset of figure (a) shows contours of the magnetic B field produced by the magnet and (b) the induced magnetic force ($f_m$) in the fluid domain. Image snapshots (i), (ii) and (iii) show the velocity field at maximum shear stress for the corresponding Weber numbers for both cases.*

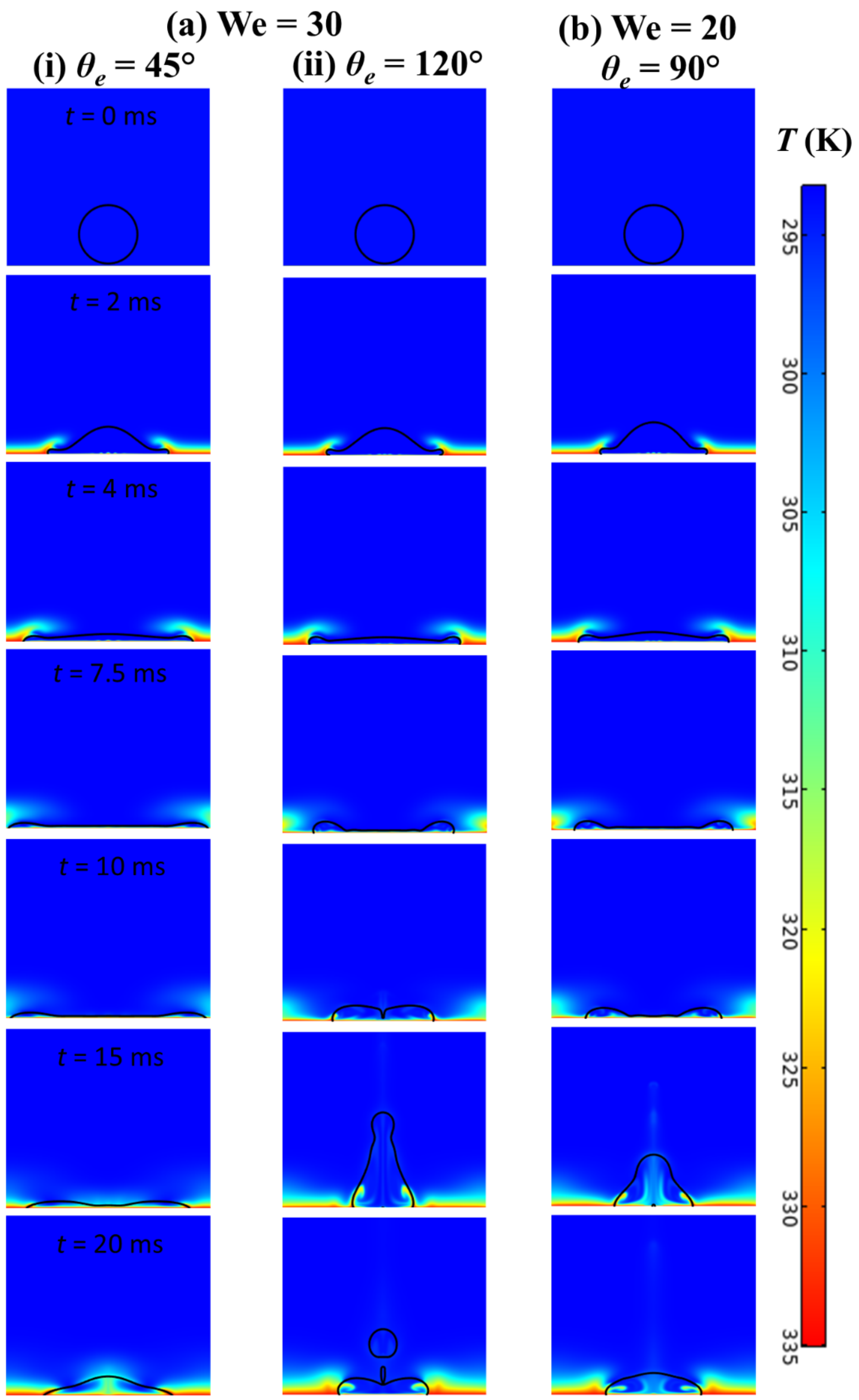

***Figure 8:*** *A comparison of morphological evolution (advancing and receding phases) of the droplet after impact at (a) We = 30 for (i) $\theta_e$ = 45° and (ii) $\theta_e$ = 120° in the presence of the magnetic field. (For We = 20, $\theta_e$ = 120° case snapshots, refer to Figure 5) (b) Evolution of the droplet after impact at We = 20 for $\theta_e$ = 90° cases in the presence of the magnetic field.*

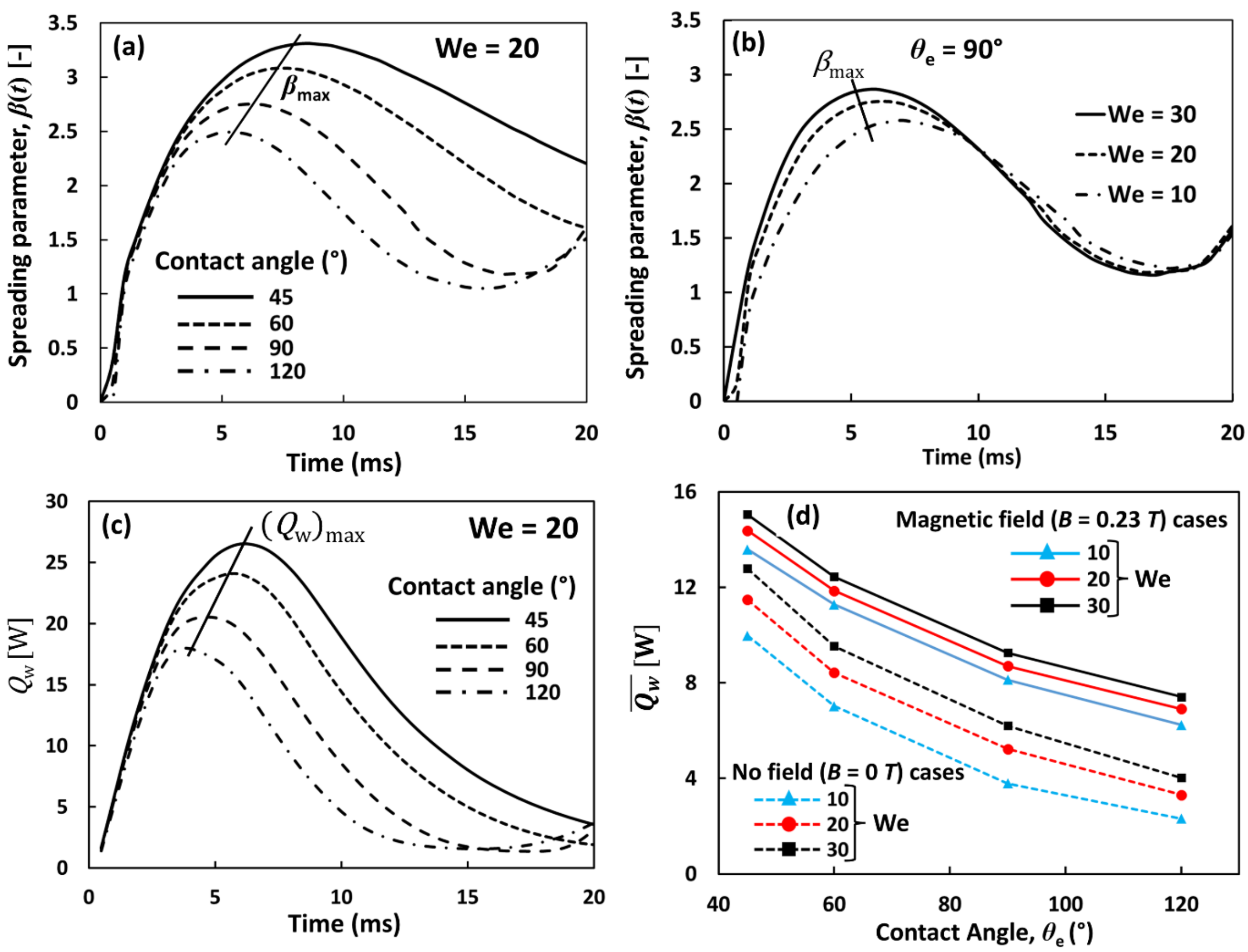

***Figure 9:*** *Comparison of plots of the spreading parameter, β(t), during evolution for (a) different wetting conditions (contact angles) at We = 20 (b) at $\theta_e$ = 90° for different Weber numbers (c) Associated temporal variation of wall heat transfer ($Q_w$) with the evolution of the impinging droplet for different contact angles at We = 20. (d) Time-averaged (20 ms) wall heat transfer ($\overline{Q_w}$) for different Weber numbers and contact angles.*

Figure 8(b), in conjunction with 9(b), illustrates the effect of Weber number on spreading dynamics for the contact angle ($\theta_e$) of 90°. Notably, varying the Weber number has minimal impact on impingement dynamics in the presence of a magnetic field. This contrasts sharply with cases without a magnetic field, where inertia plays a dominant role during the spreading phase. The induced magnetic force, approximately two orders of magnitude greater than the inertia force, effectively negates the influence of inertia and capillary forces during the spreading and receding phases. Similar observations have also been reported by Ahmed et al. [17], where they observed that while the magnitude of maximum spreading changed with applied field intensity, the maximum spreading always occurred at around 6.5 ms after impact, similar to the present case. This dominance is further evident in the time-averaged (20 ms) heat transfer data presented in Figure 9(d) for all the investigated cases. Without a magnetic field,

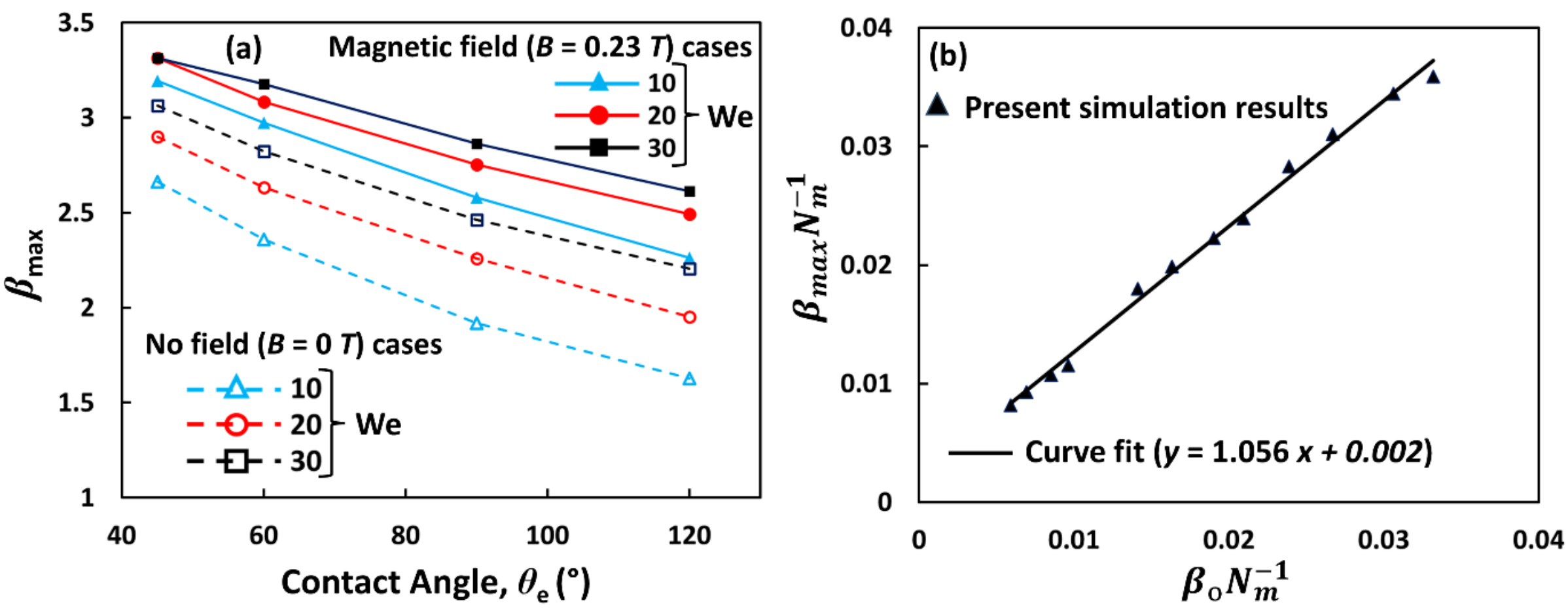

***Figure 10:*** *(a) Comparison of maximum spreading parameter ($\beta_{max}$) for both magnetic and non-magnetic field cases. (b) A general correlation is developed based on previous scaling analysis for the prediction of $\beta_{max}$ for magnetic field cases based on maximum spreading in no-field ($\beta_o$) cases.*

the expected trends are observed: heat transfer increases with increasing inertia force (Weber number) and decreases with increasing contact angle (as the surface becomes more hydrophobic). However, with the magnetic field applied, the differences between cases diminish significantly, indicating that the magnetic field's influence overshadows the contributions of inertia and capillary forces. A maximum enhancement of about 170% for the We = 10 and ($\theta_e$) = 120° case and a minimum enhancement of ~ 17% for the We = 30 and ($\theta_e$) = 45° case are observed in the presence of the magnetic field among all the investigated cases. The enhancement in heat transfer decreases with increasing Weber number, as seen in Figure 9(d). A general-purpose correlation is developed to predict the maximum spreading diameter ($\beta_{max}$) in an applied non-uniform magnetic field (which can give rise to the Kelvin body force). In the present work, both the Weber number and the contact angle are changed while other parameters are kept constant. The Weber number is changed by changing the inertia force, while surface tension forces are kept constant. The present findings align closely with those reported by Benther et al. [23] from their analogous experimental study. Specifically, they observed an average heat transfer rate of ~ 14 watts under magnetic field conditions, at We = 60 and $\theta_e$ = 45°, which corroborates the heat transfer magnitudes obtained in this current investigation, thereby lending further credence to the reliability and accuracy of the computational methodology employed herein. Figure 10(a) presents a comparison of the maximum spreading parameter ($\beta_{max}$) for both magnetic field and non-magnetic field cases of ferrofluid. The maximum change in $\beta_{max}$ is observed for We = 10 cases, and the effect of the applied magnetic field is increasingly overcome by the inertial forces, as the Weber number increases to 30. The maximum observed change is around 40% at We = 10, $\theta_e = 120°$. This indicates that a tuning of the magnetic force is required to get the desired effect at higher Weber numbers.

Finally, a general relationship is proposed to predict the maximum spreading ($\beta_{max}$) in the presence of a magnetic field, based on scaling arguments presented by Li et al. [21] and Huang et al. [22]. It is known that the initial droplet spreading is driven by an increase in the local pressure at the centre of the droplet upon impact. This increase in pressure causes droplets to spread. As droplets spread, capillary and viscous forces progressively increase and finally overcome inertial forces at maximum spreading. In the pressure-driven stage, the time scale of droplet spread is given by $t \sim d_o/U$. When a magnetic field is applied, the induced magnetic force, which is attractive in nature and oriented in the same direction as the inertia forces at the time of impact, makes the pressure gradient stronger. They argued that the induced magnetic force aids the inertia force during the droplet spreading phase, and the motion of the droplet is governed by the magnetic force. Through scaling arguments, Li et al. proposed a non-dimensional magnetic parameter, $N_m \sim \mu_o \rho U^2/B^2$, defined as the ratio of inertia to magnetic force to describe the motion of the droplet in the pressure-driven regime. Also, by assuming the complete conversion of kinetic energy to surface energy and dissipation into thermal energy at maximum spreading, their scaling analysis relates maximum spreading in the presence of a magnetic field with magnetic parameter ($N_m$) as $\beta_{max} \sim N_m$. Based on their analysis, Li et al proposed a correlation to predict maximum spreading for the magnetic field cases by relating it to maximum spreading with no magnetic field case at the same Weber number and $N_m$, as $\beta_{max} \sim f(\beta_o N_m^{-1})$, where $\beta_o$ represents maximum spreading for no field cases. The proposed functional form for the correlation is $\beta_{max} N_m^{-1} = P/(1 + AP)$ where $P = \beta_o N_m^{-1}$ and $A$ is the curve fitting coefficient. By relating $\beta_{max}$ with $\beta_o$, they accounted for viscous and capillary effects in spreading dynamics. Huang et al. used a similar scaling approach to relate $\beta_{max}$ with $N_m$ as $\beta_{max} \sim N_m^{-1/2}$. They also proposed a similar correlation to Li et al. for the prediction of the maximum spreading parameter in the presence of a magnetic field.

Based on earlier scaling analysis, a general relationship is also proposed in the present work to predict the maximum spreading parameter in the presence of the magnetic field. Figure 10(b) shows a curve fit to the present data. A linear function of the form, $y = a\,x + b$, has been used to fit the outcome of the present work. Here, $y = \beta_{max} N_m^{-1}$ and $x = \beta_o N_m^{-1}$ represent the same physical parameters as in previous works. $a$ and $b$ are the curve slope and y-axis intercept, respectively, with values of 1.056 and 0.002 for the present work. To validate this functional form, the experimental results of Banthar et al. [23] have been used. The $\beta_{max}$ value for the heated substrate, magnetic field case, is around 2.50 in their case, and the proposed correlation predicts it to be around 2.45, which is close to the actual experimental values. This further reinforces confidence in the proposed correlation.

In summary, when the induced magnetic force is significantly stronger than inertia, capillary, and viscous forces, it becomes the primary driver of the impingement dynamics and associated heat transfer characteristics.

## Summary and Conclusions:

This study examined the thermal transport behavior of ferrofluid droplets impinging on heated surfaces under the influence of a magnetic field. The magnetic field, generated by a finite-sized permanent magnet, was modelled using Maxwell's equations. Droplet dynamics and interfacial evolution on both hydrophilic and hydrophobic surfaces were captured using a unified set of mass, momentum, and energy equations, coupled with a phase-field formulation.

In the absence of a magnetic field, surface wettability played a significant role in the spreading and receding phases; hydrophilic surfaces enhanced spreading, while hydrophobic surfaces promoted recoil. However, the application of a magnetic field introduced an attractive magnetic body force due to the ferrofluid's super-paramagnetic nature. This force notably increased the maximum spreading diameter on both surface types and suppressed rebound on hydrophobic surfaces. The magnetic force effectively augmented the inertial force and counteracted capillary effects, thereby altering the spreading dynamics.

Additionally, the induced magnetic forces modified the internal velocity field, resulting in a pronounced wall shear effect. For instance, at a Weber number of 10 and a contact angle of 120°, the average wall shear stress increased by approximately 200%. Correspondingly, a maximum enhancement of about 75% in average heat transfer was observed at a Weber number of 30 and a contact angle of 120°, attributed to the increased contact area and reduced rebound.

Overall, the findings demonstrate that magnetic fields significantly influence droplet spreading, wall shear stress, and heat transfer rates, particularly on hydrophobic surfaces. While the impact is less pronounced on hydrophilic surfaces, it remains substantial, especially at lower Weber numbers.

The ability to actively control droplet behavior and thermal performance using external magnetic fields presents promising opportunities for dynamic thermal management. This approach could be particularly beneficial in systems with fluctuating heat loads, such as in the design of thermal switches or adaptive cooling technologies.

## ACKNOWLEDGEMENT

The authors would like to thank UPES and Amity University for providing the required facilities and resources to conduct this work. The authors would also like to thank Dr Debartha Chatteerjee, IIT Kanpur, for his support in performing computer simulations and Dr Ankush Jaiswal, IIT Hyderabad, for reviewing and providing valuable suggestions for this work. This research did not receive any specific grant from funding agencies in the public, commercial, or not-for-profit sectors.

## AUTHOR DECLARATIONS

### Conflict of Interest

The authors have no conflict of interest to disclose.

### Data Availability

The data that support the findings of this study are available within the article.